\documentclass{PoS}

\newcommand{\ie}{{\em i.e.}}
\newcommand{\eg}{{\em e.g.}}

\newcommand{\ex}[1]{\cdot 10^{#1}} 

\usepackage{amssymb}
\usepackage{amsmath}
\usepackage{graphicx}
\usepackage{cite}
\usepackage{bm}
\usepackage{epsfig}
\usepackage{multirow}

\title{Classifying extremely imbalanced data sets}

\ShortTitle{Classifying extremely imbalanced data sets}

\author{\speaker{Markward Britsch}\\
        Max-Planck-Institut f\"ur Kernphysik, PO Box
	103980, 69029 Heidelberg, Germany \\
        E-mail: \email{markward@mpi-hd.mpg.de}}

\author{Nikolai Gagunashvili\\
        University of Akureyri, Borgir, v/Nordursl\'od, IS-600 Akureyri, Iceland\\
        E-mail: \email{nikolai@unak.is}}

\author{Michael Schmelling\\
        Max-Planck-Institut f\"ur Kernphysik, PO Box
	103980, 69029 Heidelberg, Germany \\
        E-mail: \email{Michael.Schmelling@mpi-hd.mpg.de}}

\abstract{Imbalanced data sets containing much more background than signal
instances are very common in particle physics, and will also be
characteristic for the upcoming analyses of LHC data. Following
up the work presented at ACAT 2008, we use the multivariate
technique presented there (a rule growing algorithm with the
meta-methods bagging and instance weighting) on much more
imbalanced data sets, especially a selection of D0 decays
without the use of particle identification. It turns out that
the quality of the result strongly depends on the number of
background instances used for training. We discuss methods to exploit this in
order to improve the results significantly, and how to handle and reduce the
size of large training sets without loss of result quality in general. We will
also comment on how to take into account statistical fluctuation in receiver
operation characteristic curves (ROC) for comparing classifier methods.}

\FullConference{13th International Workshop on Advanced Computing and Analysis Techniques in Physics Research, ACAT2010\\
		February 22-27, 2010\\
		Jaipur, India}

\begin{document}

\section{Introduction}
  Multivariate analysis has successfully been employed in many high
  energy physics data analyses, see, \eg, \cite{higgs, babar2, babar1}. Of particular interest is the common
  case in which the background  dominates the signal. 
%In the data mining community
%  (in computer science multivariate analysis is also
%  called ``data mining''), such problems where there are, \eg,
%  many more background than signal events, are referred to as
%  ``imbalanced problems'' (see, \eg, \cite{weiss}). 
In intelligent data processing, such problems where there are, \eg, many more background
than signal events, are referred to as ``imbalanced data problems ''(see, \eg, \cite{weiss}).

At ACAT 2008 we have presented a method for imbalanced problems consisting of
three components for classifying imbalanced data sets~\cite{acat08}. It has been
tested on a $\Lambda$ selection with a background to signal ratio of less than 100. Here we test the same method on a $D^0$-selection without the usage of particle
identification with Monte Carlo data produced for the LHCb experiment~\cite{LHCb}. This data has a background to signal ratio of about
3000 and is thus much more imbalanced. It turns out that this extreme
imbalance needs special care which we will describe here in detail. The result of this selection has already been presented at DIS 2009~\cite{dis09} and shown
to be superior to a cuts based analysis. 
Since the classification
method has already been presented at ACAT 2008, it will be summarized only briefly in the following.  

The first of the three components of our method is RIPPER~\cite{cohen, tan, witten}, a rule based learner. Often a classifier gives a discriminant (like the probability for a candidate
to be signal) as an output. This is used by choosing a cut value on this variable to adjust to
the signal to background ratio in the data set and to one's needs. Instead, RIPPER, as it is used here, only gives a binary output, \ie, classifying the candidate to be signal or background. We use a cost based method as the second component of our method. The way we are using the cost is by introducing
weights in the {\em training} step. This is called instance weighting and it follows that
we get a new classifier model for each choice of cost~\cite{ting, witten}. The reason is
that in many cases the model building uses the error rate to decide on the rules
or tree branches. But the error rate depends on the signal to background ratio
in the sample which is changed due to the weights. Instance weighting provides
more effective and simple models for classifiers like decision trees or rule
based learners~\cite{zhao}. Our third component is bagging (bootstrap aggregation)~\cite{breiman} which is used to
stabilize the algorithm. 
%It works like boosting, but without the usage of weights.
It works like boosting, but without the usage of weights and  does not lead to
over fitting.
For large training sets we introduce one or two preselection steps to prevent memory overflow
and to reduce the training time.

For implementing the classification method we are using the well known data mining
package WEKA~\cite{witten, weka}. WEKA is a free software written in java that implements many ready to use data mining algorithms like supervised and unsupervised classification. It can be used via a graphical user interface or by the command line.
Our sequence consists of the following steps:
bagging, set the costs for instance weighting and applying the RIPPER classifier. For
each preselection an extra full classification step is done including
bagging. The costs can be represented in a cost matrix like those in Tables
\ref{t:costPresel} and \ref{t:costMain}. Each entry in such a matrix is the cost to be used in training depending on whether the instance is a true signal or background (row) and on the prediction of the classifier (column).
For preselections we put a high cost
for loosing $D^0$ to keep almost all of them while reducing the background
significantly (see Table \ref{t:costPresel}). In the main classification step we
then use a high cost $x$ for wrongly accepted background as shown in Table
\ref{t:costMain}. To produce the ROC curve we scan the cost parameter $x$, so we
have one classifier model per point in the ROC curve. 

 \begin{table}
 	\centering
\begin{minipage}[t]{0.48\textwidth}
  \centering
 \begin{tabular}{c|c|c}
     & predicted BG & predicted $D^0$ \\
    \hline
    true BG & 0 & 1 \\
    \hline
    true $D^0$ & 200 & 0\\
  \end{tabular}
  \caption[A sample preselection cost matrix.]{\label{t:costPresel} A sample
    cost matrix for preselection. The number 200 varies with the number of preselections.
%Here we have 0 cost for correctly classified candidates, 1 for background
%    predicted to be signal and a cost of 200 for true signal classified as
%    background.
}
%\end{table}
\end{minipage}
\begin{minipage}[t]{0.48\textwidth}
%\begin{table}
  \centering
  \begin{tabular}{c|c|c}
     & predicted BG & predicted $D^0$ \\
    \hline
   true BG & 0 & $x$ \\
    \hline
   true $D^0$ & 1 & 0 \\
  \end{tabular} 
  \caption[The cost matrix for the main selection.]{\label{t:costMain} The cost
    matrix for the main selection.}
  \end{minipage}
\end{table}

\section{$D^0$-meson selection in LHCb Monte Carlo}

LHCb is one of the four large experiments at the $pp$-collider
LHC. It is built for precision measurements of $CP$ violation and rare
decays and is designed as a forward spectrometer.

To select $D^0$-mesons, we use the decay $D^0 \rightarrow \pi^+K^-$. The
data we are using is minimum bias Monte Carlo, $3.6 \ex{7}$ events produced in
2006 for the LHCb experiment at a center of mass energy of $\sqrt{s} = 14$
TeV. Candidates are pairs of oppositely charged tracks passing through the full
spectrometer, with the application of a very loose preselection cut on the
distance of closest approach ($DoCA$) of the two tracks of $DoCA < 10$~mm. We
use 14 geometric, track quality and kinematic variables. The training data sets
contain the same number of signal but increasing number of background candidate (see Table
\ref{t:setsD0}). 

\begin{figure}
  \centering \vspace*{-0.0 cm} \hspace*{-0.0 cm}
  \includegraphics[angle=-90, width=0.6\textwidth]{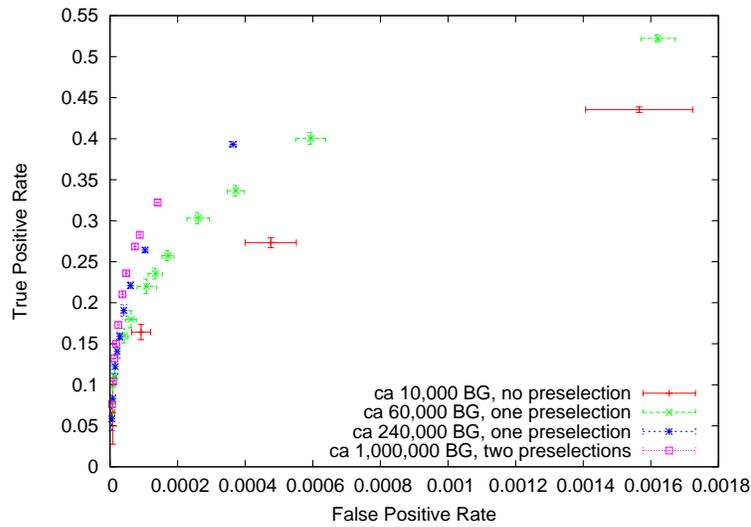}
  \vspace*{0.7 cm} 
  \caption[ROC curves for $D^0$ selections.]{\label{f:D0roc} ROC curve, \ie, true positive rate (signal efficiency) versus false positive rate (background efficiency), for using the different training samples. Mind that in this representation a curve being more to the upper left is better.}
\end{figure}

\begin{figure}
  \centering \vspace*{-0.0 cm} \hspace*{-0.0 cm}
  \includegraphics[angle=-90, width=0.6\textwidth]{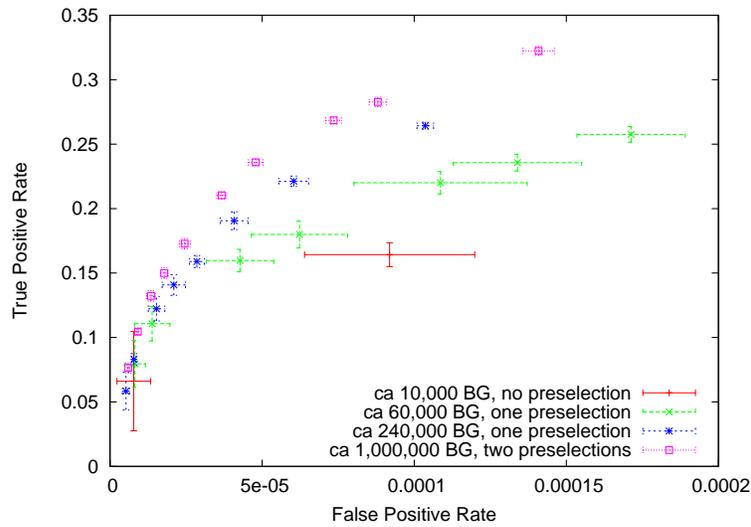}
  \vspace*{0.7 cm} 
  \caption[ROC curves for $D^0$ selections, zoom.]{\label{f:D0rocZoom} As Figure \ref{f:D0roc}, but zoomed in.}
\end{figure}

\begin{figure}
  \centering \vspace*{-0.0 cm} \hspace*{-0.0 cm}
  \includegraphics[angle=-90, width=0.6\textwidth]{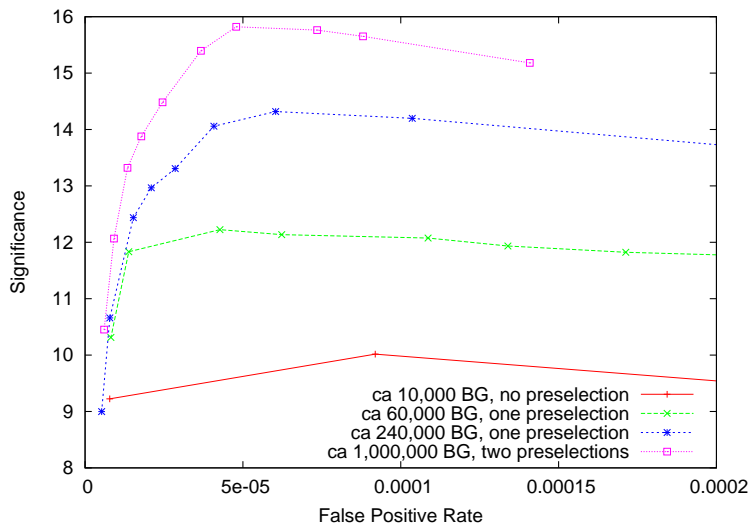}
  \vspace*{0.7 cm} 
  \caption[Significance curves for $D^0$ selections.]{\label{f:D0sign} Significance versus false positive rate for the $D^0$ selections using the different training samples.}
\end{figure}

In Figure \ref{f:D0roc} the receiver operation characteristic (ROC) curves for
using the four different training sets are shown (the plots are done using the
test set). The ROC curve is defined as a plot of the true positive rate (or
signal efficiency) versus the false positive rate (or background
efficiency). We find that those classifier models corresponding to the training
sets with larger background give superior results with respect to those where a
training set with lower background has been used. This is especially evident in
a zoom in Figure \ref{f:D0rocZoom}. Here, as almost everywhere else, we
see that for a false positive rate of around $5\ex{-5}$ the classifier model
corresponding to the largest background in the training sample is the best. From
Figure \ref{f:D0sign} we see that this region in false positive rate is where
the highest significance\footnote{Significance is defined here as $\frac{\rm \#signal}{\sqrt{\rm \#background
    + \#signal}}$} is. Thus this is an important working point. 

\begin{table}
  \centering
  \begin{tabular}{c|c|c|c}
    data set & ca \# BG & \# sig. & \# presel. \\
    \hline
    test & $6.5\ex{6}$ & 1827 & --\\
    \hline
    training small & 10,000 & 1851 & 0\\
    training mid & 60,000 & 1851 & 1 \\
    training larger & 240,000 & 1851 & 1 \\
    training largest & 1,000,000 & 1851 & 2 \\
  \end{tabular}
  \caption[The $D^0$ training and testing data sets.]{\label{t:setsD0}The $D^0$
  training and testing data sets. The second column contains the number of background
  candidates, the third column contains the number of signal candidates and the last
  column gives the number of preselections used.}
\end{table}

\begin{figure}
%\centering
\begin{minipage}[t]{0.48\textwidth}
  	\centering 
	\includegraphics[width = 0.5\textwidth]{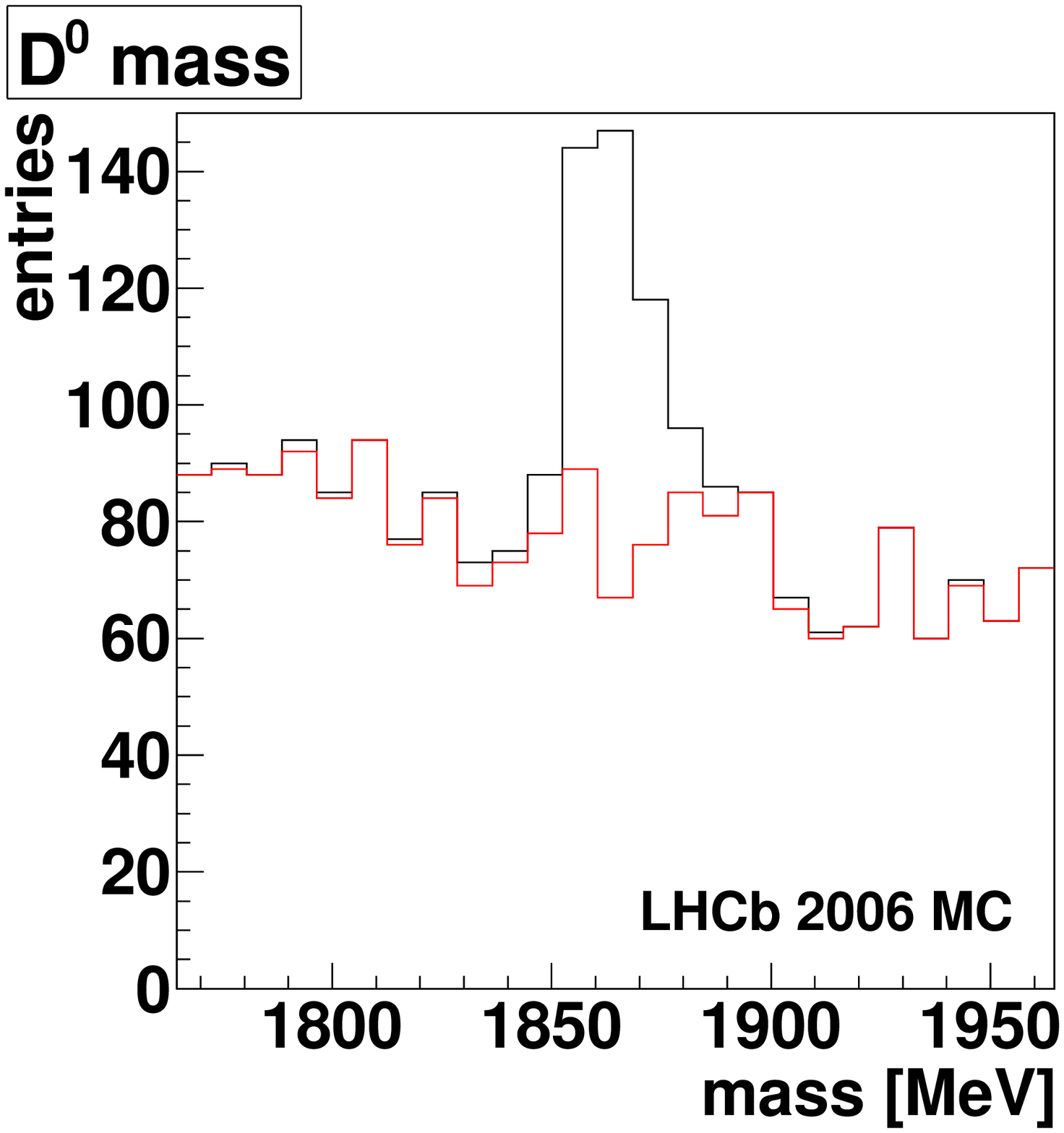} 
	\caption[Mass plot cuts based.]{\label{f:massCuts} The $D^0$ mass plot
	  after a cuts based selection using the same variables.} 
%\end{figure}
\end{minipage}
\begin{minipage}[t]{0.48\textwidth}
%\begin{figure}
  	\centering 
	\includegraphics[width = 0.5\textwidth]{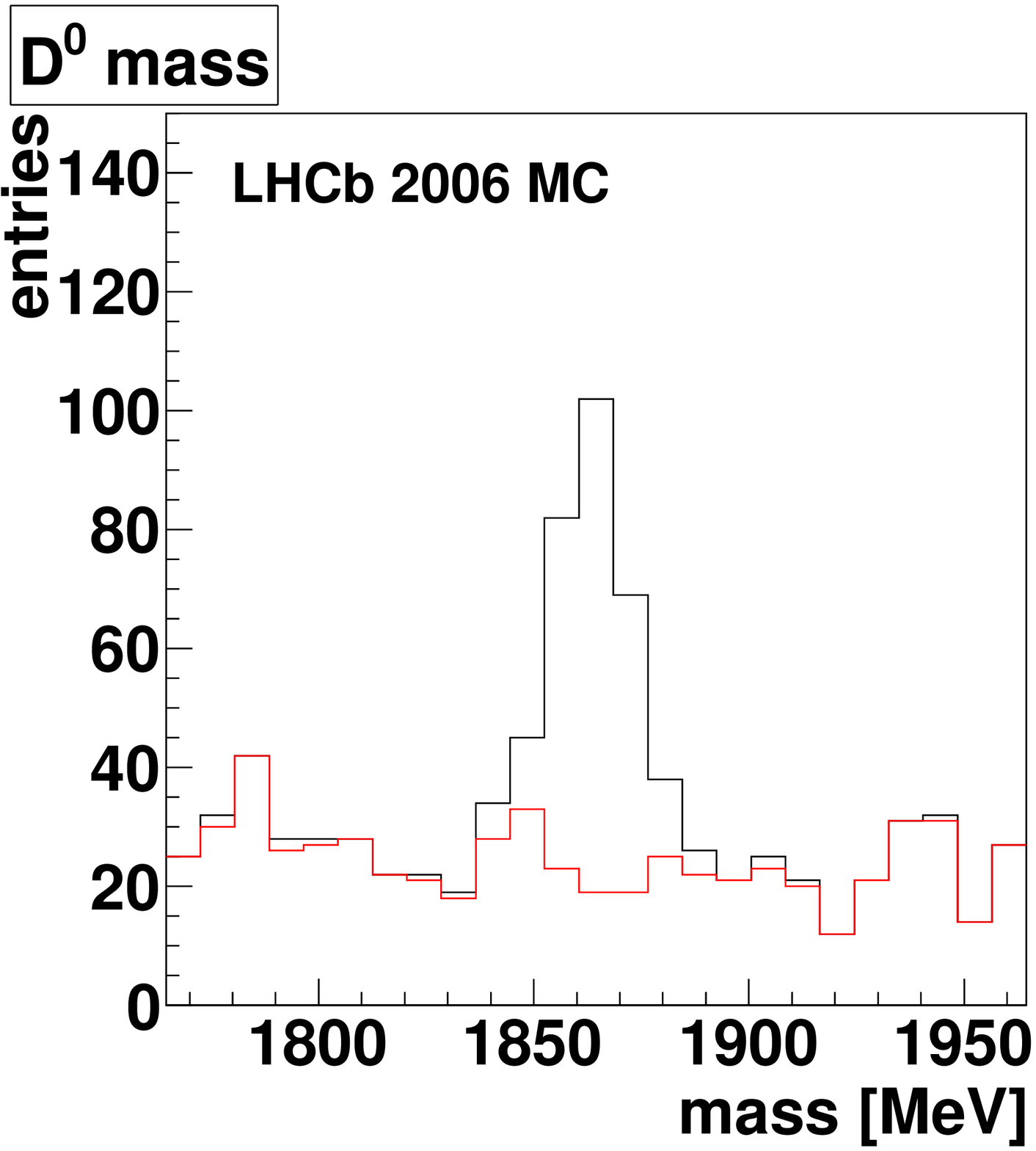}
	\caption[Mass plot MVA selection]{\label{f:massMVA} The mass plot after
	  our multivariate method, cost parameter set in a way to get the same signal yield as the
	  cuts based analysis for comparison reasons.} 
\end{minipage}
\end{figure}

Figures \ref{f:massCuts} and \ref{f:massMVA} compare the mass plots of a cuts
based $D^0$ selection using the same variables, and this multivariate method
where the cost has been chosen to get the same signal yield as in the cuts based
scheme. This was done for comparison reasons and we see that for the same
signal yield the background is reduced drastically. 
 
\section{Forest cover type data}

Is this behavior special to our data set or does it also appear on
other kinds of data? From a data mining data set
repository~\cite{repository},  we choose the data set called forest cover type (see
also~\cite{covType}). It is about predicting forest cover type from cartographic
variables. The
observations (30 $\times$ 30 meter cells) are wilderness areas with minimal human-caused disturbances as determined by the US Forest
Service (USFS) in the Roosevelt National Forest of northern Colorado. The 54 variables include 10 integer variables, like elevation in feet, slope in degrees and vertical distance to nearest surface water. The rest of the variables are of categoric type indicating the wilderness and soil type.
The classes to predict are seven cover types, like Spruce/Fir, Lodgepole Pine or Ponderosa Pine.
We use the
10 integer variables only and use class 4 (Cottonwood/Willow) as
``signal'', the rest as ``background'' to get an imbalanced data set. Splitting up the data set into test and training data, we have about 290,000 background instances and 1365 signal instances in the test set. For training about 240,000 background and 1382 signal instances are left. 
Again we use different training sets with the same number of signal (1382) but increasing
number of background, namely 10,000, 60,000, 240,000 and 5 $\times$ 240,000, where in the last case we use a method to artificially replicate the background instances as described below.
Also the number of preselections increases with the number of background instances in the training. We use no preselection in the case of 10,000 background instances, one preselection in the cases of 60,000 and 240,000 background instances and two preselections for the larges training sample.
% as shown in Table \ref{t:setsCov}. 
In this larges training sample we use additional artificial
background data obtained by four times randomization of existing
background instances using the SMOTE algorithm~\cite{chavla}. This was done to see if we can
improve the result in spite of the fact that no more background events have been
available. 

\begin{figure}
	\centering
	\begin{minipage}[t]{0.45\textwidth}
  \centering \vspace*{-0.0 cm} \hspace*{-0.0 cm}
  \includegraphics[angle=-90, width=0.9\textwidth]{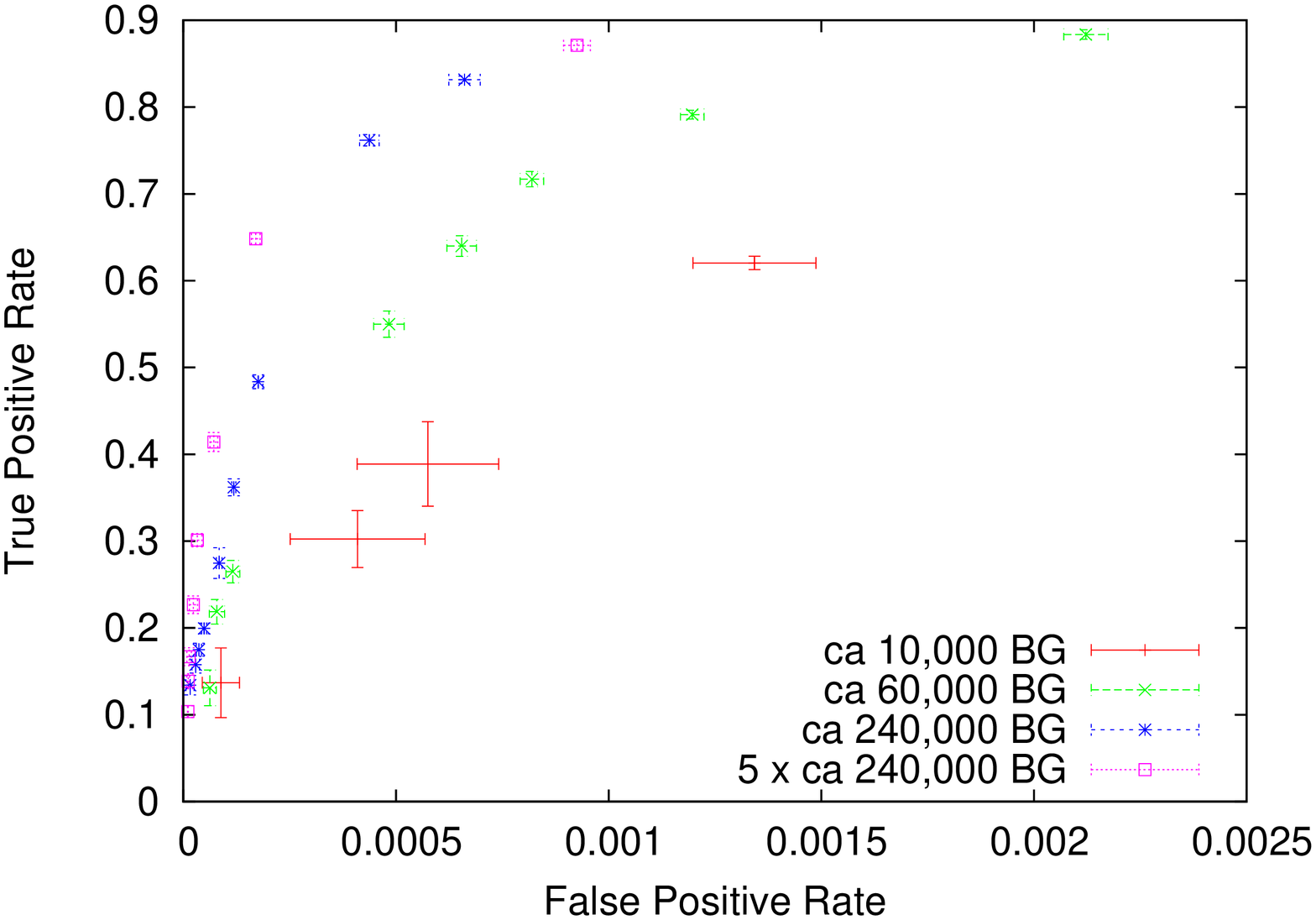}
  \vspace*{0.7 cm} 
  \caption[ROC curve for forest cover type data.]{\label{f:CovRoc} The ROC curve for using the different training samples for the cover type data set.}
\end{minipage}
%\begin{figure}
\hspace{0.01\textwidth}
	\begin{minipage}[t]{0.45\textwidth}
  \centering \vspace*{-0.0 cm} \hspace*{-0.0 cm}
  \includegraphics[angle=-90, width=0.9\textwidth]{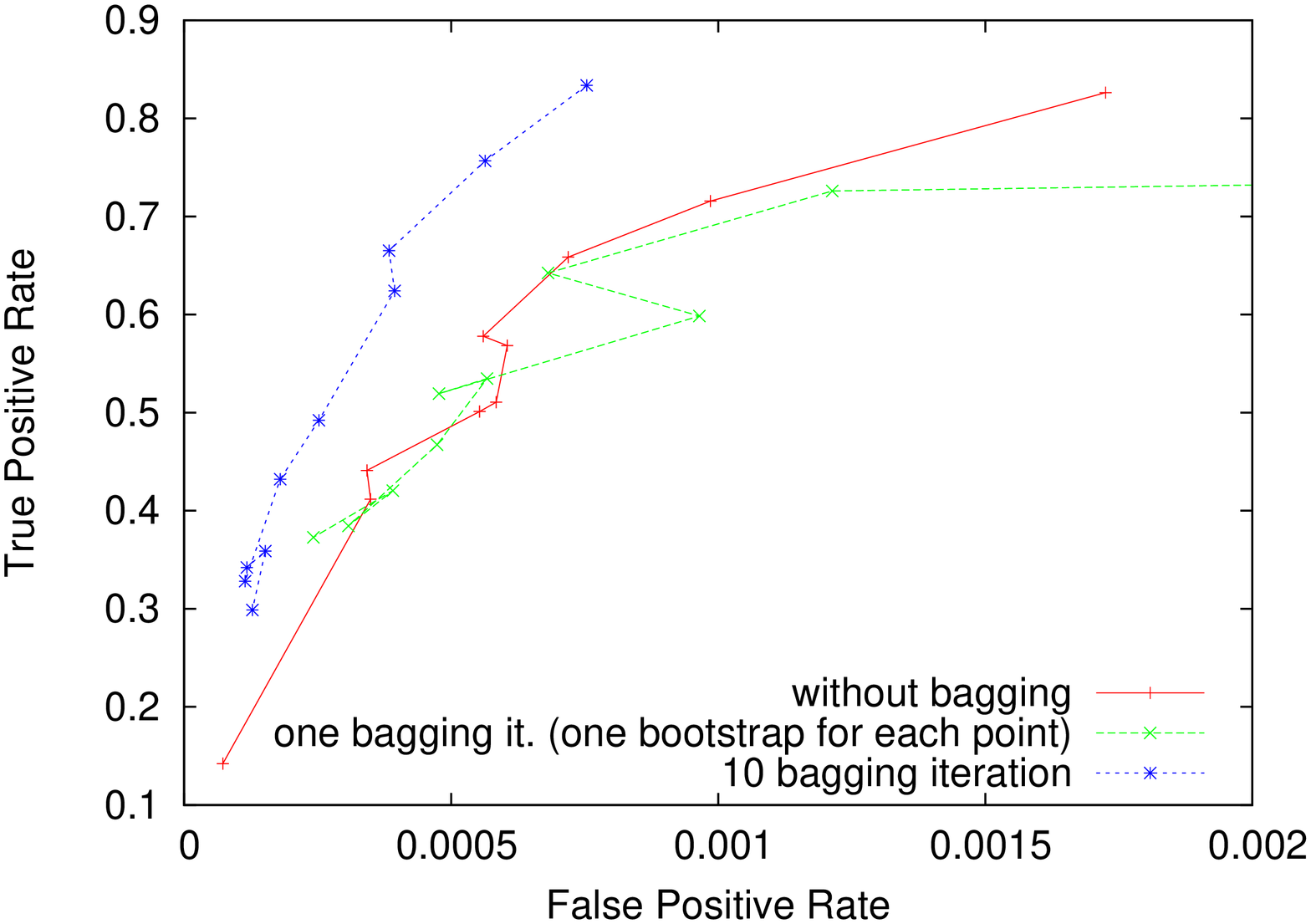}
  \vspace*{0.7 cm} 
  \caption[Scatter due to differences in the training data.]{\label{f:singleRoc}
    The red ROC curve is plotted using no bagging, \ie, each point has been done
    using the same training sample. For the green curve, for each point the
    training set has been re-sampled, \ie, using one bagging iteration including a
    change in the random seed for each point in ROC space. The blue curve shows
    the effect of many (10) bagging iterations.} 
\end{minipage}
\end{figure}

In Figure \ref{f:CovRoc} we present the corresponding ROC curves. Again we see
the same effect as for the $D^0$ data. In addition we see that 
adding artificial background data also improves the result.

%\begin{table}
%  \centering
%    \begin{tabular}{c|c|c|c}
%      data set & ca \# BG & \# sig. & \# presel. \\
%      \hline
%      test & 290,000 & 1365 & --\\
%      \hline
%      training small & 10,000 & 1382 & 0\\
%      training mid & 60,000 & 1382 & 1 \\
%      training large & 240,000 & 1382 & 1 \\
%      training artificial & 5 $\times$ 240,000 & 1382 & 2 \\
%    \end{tabular}
%  \caption[The cover type training and testing data sets.]{\label{t:setsCov} The
%  same as in Table \ref{t:setsD0}, but for the forest cover type
%  training and testing data sets. For an explanation of the
%  artificial data see text.}
%\end{table}
%

%\begin{figure}
%  \centering \vspace*{-0.0 cm} \hspace*{-0.0 cm}
%  \includegraphics[angle=-90, width=0.6\textwidth]{gnuplot_acatCov-2.eps}
%  \vspace*{0.7 cm} 
%  \caption[ROC curve for forest cover type data.]{\label{f:CovRoc} The ROC curve for using the different training samples for the cover type data set.}
%\end{figure}

\section{How to compare ROC curves with scatter}

We have a different classifier model for each point in ROC space. But these
classifier models depend not only on the training sample choice, but they also
depend on random choices in bagging and in RIPPER during the training. Thus the
pure ROC curves look noisy. 
%So we need a way to smooth the curve (\ie, average
%many) and a measure for the scatter (\ie, error bars). 
So we need a way to find the expectation curve (\ie, average many) and a measure for
the scatter (\ie, error bars). 

%\begin{figure}
%  \centering \vspace*{-0.0 cm} \hspace*{-0.0 cm}
%  \includegraphics[angle=-90, width=0.6\textwidth]{gnuplot_single_vgl2-2.eps}
%  \vspace*{0.7 cm} 
%  \caption[Scatter due to differences in the training data.]{\label{f:singleRoc}
%    The red ROC curve is plotted using no bagging, \ie, each point has been done
%    using the same training sample. For the green curve, for each point the
%    training set has been re-sampled, \ie, using one bagging iteration including a
%    change in the random seed for each point in ROC space. The blue curve shows
%    the effect of many (10) bagging iterations.} 
%\end{figure}

In Figure \ref{f:singleRoc}, the red curve uses the {\bf same} sample for
training for all points, for the green curve the training set has been re-sampled for
each point. The less noisy curve (red) hides its scatter, \ie, its dependence on
the training set. The same is true for ordinary ROC curves
  using a cut on a discriminant. The more noisy curve (green) tells us
something about this scatter. As it should be, bagging reduces this scatter by using
many bagging iterations (blue curve).

There are different methods for averaging ROC curves and to get error bars
discussed in literature (see,
\eg,~\cite{accEstComparing,robustClassification,roc,
  rocBandsMeth,pointwRocConf,rocBandsEmpirical }). But  none (that we could find) takes into
account the scatter due to the training set. 
%So the following is our rather ad
%hoc method. 
In our method we start by doing each main selection 10 times with different random
seeds. Then we take the mean false positive rate (FPR) and true positive rate (TPR)
as the point in ROC space. This is similar to using 10 cross-validation samples
used in literature. But now we take the standard deviations as errors
in FPR and TPF. The result is what is shown in the plots in Figures \ref{f:D0roc}, \ref{f:D0rocZoom} and \ref{f:CovRoc}. What is the
distribution like?

\begin{figure}
	\centering
	\begin{minipage}[t]{0.34\textwidth}
  \centering \vspace*{-0.0 cm} \hspace*{-0.0 cm}
  \includegraphics[width=1.0\textwidth]{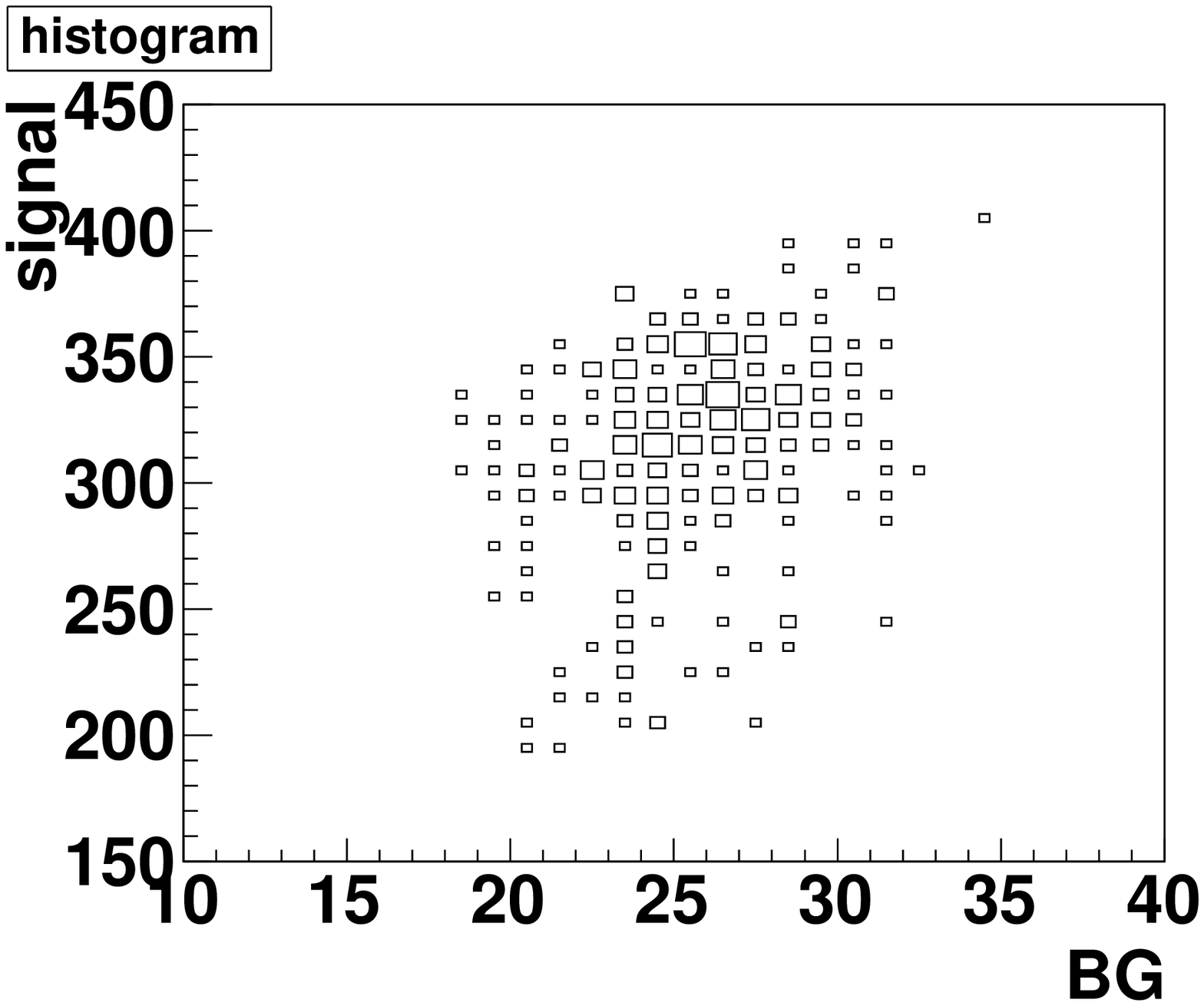}
  \vspace*{-0.0 cm} 
%  \caption[Scatter plot, 300 points with the same cost.]{\label{f:sBox} A scatter (number of signal versus number of background) plot of 300 points using the same cost but different random seeds.}
%\end{figure}
\end{minipage}
%\begin{figure}
	\begin{minipage}[t]{0.32\textwidth}
  \centering \vspace*{-0.0 cm} \hspace*{-0.0 cm}
  \includegraphics[width=1.0\textwidth]{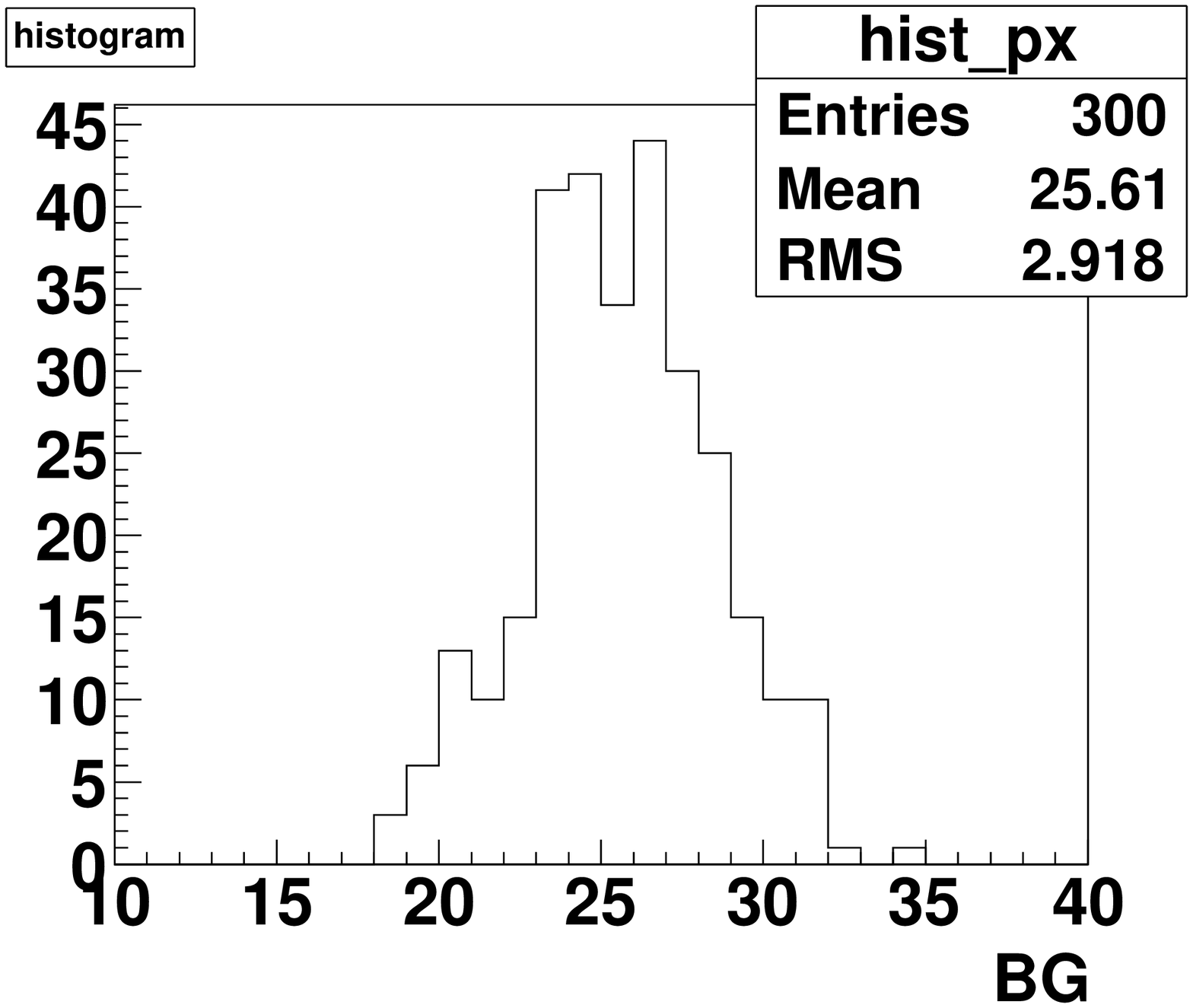}
  \vspace*{-0.0 cm} 
%  \caption[Background distribution for 300 points using the same cost.]{\label{f:sBG} The number of background for 300 points using the same cost but different random seeds.}
%\end{figure}
\end{minipage}
%\begin{figure}
	\begin{minipage}[t]{0.32\textwidth}
  \centering \vspace*{-0.0 cm} \hspace*{-0.0 cm}
  \includegraphics[width=1.0\textwidth]{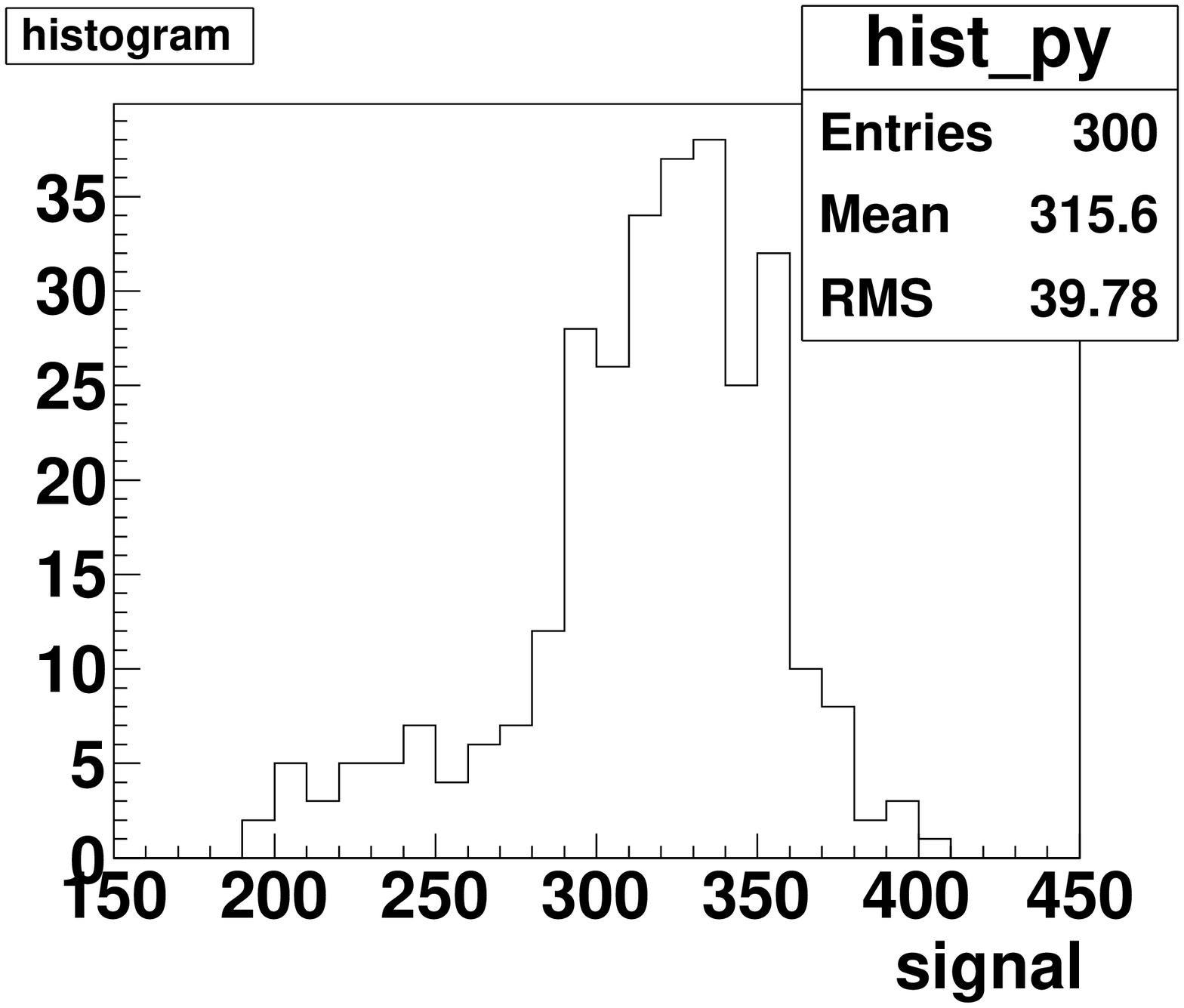}
  \vspace*{-0.0 cm} 
%  \caption[Signal distribution for 300 points using the same cost.]{\label{f:sSig} The number of signal for 300 points using the same cost but different random seeds.}
\end{minipage}
  \caption[300 points using the same cost.]{\label{f:scatter} Shown on the left is a scatter  plot (number of signal versus number of background) of 300 points using the same cost but different random seeds. In the middle its projection on the background axis is shown and on the right the projection on the signal axis.}
\end{figure}

To find this out, we are using 300 samples of the same cost but different random
seeds -- with no averaging. This distribution in number of signal versus number of
background candidates is shown in Figure \ref{f:scatter} including the projections onto the background and signal axis respectively. 
%The distributions are asymmetric and have tails, thus the standard
%deviation can not easily be interpreted as a confidence level.
%The distributions are asymmetric and have tails, thus the standard deviation can
%not easily be used for defining a confidence level.
  The distributions are asymmetric and have tails, thus the
  standard deviation cannot be associated with a well defined
  confidence level.
  Nevertheless if we calculate the 68~\% confidence level intervals for the background and the signal histograms in Figure~\ref{f:scatter}, we get [23, 28] and [282, 351] respectively. This, possibly by pure coincidence, is very close to what we get as the one standard deviation interval from the mean and the RMS, \ie, [22.7, 28.5] for background and [276, 356] for signal respectively.
  %, even though the actual example suggests
%  that the probability content of $\pm$ 1 standard deviation is close to 68~\%.
% dogma

%\begin{figure}
%	\centering
%	\begin{minipage}[t]{0.3\textwidth}
%  \centering \vspace*{-0.0 cm} \hspace*{-0.0 cm}
%  \includegraphics[width=1.0\textwidth]{scatterBox-2.eps}
%  \vspace*{-0.0 cm} 
%  \caption[Scatter plot, 300 points with the same cost.]{\label{f:sBox} A scatter (number of signal versus number of background) plot of 300 points using the same cost but different random seeds.}
%%\end{figure}
%\end{minipage}
%%\begin{figure}
%	\begin{minipage}[t]{0.3\textwidth}
%  \centering \vspace*{-0.0 cm} \hspace*{-0.0 cm}
%  \includegraphics[width=1.0\textwidth]{scatterBG-2.eps}
%  \vspace*{-0.0 cm} 
%  \caption[Background distribution for 300 points using the same cost.]{\label{f:sBG} The number of background for 300 points using the same cost but different random seeds.}
%%\end{figure}
%\end{minipage}
%%\begin{figure}
%	\begin{minipage}[t]{0.3\textwidth}
%  \centering \vspace*{-0.0 cm} \hspace*{-0.0 cm}
%  \includegraphics[width=1.0\textwidth]{scatterSig-2.eps}
%  \vspace*{-0.0 cm} 
%  \caption[Signal distribution for 300 points using the same cost.]{\label{f:sSig} The number of signal for 300 points using the same cost but different random seeds.}
%\end{minipage}
%\end{figure}

\section{Conclusion and Outlook}

For extremely imbalanced data sets we have seen that more background in the training
set is better for the LHCb $D^0$ selection as well as the forest cover type
data set -- in an important region of false positive rate. One or two
preselections with less background helps reducing the data to handle large
training sets. Even using extra artificial background instances helps.

For ROC curve errors, we have presented a method which seems
reasonable and practical but the error-bars cannot be interpreted as a confidence level. 

More sophisticated ways to reduce the data size without loosing classification
quality have also been investigated by the authors~\cite{newBritsch}. Future work will include to search for better
ways to average ROC curves and to produce error bars. In addition we want to try
different classifiers (e.g., decision trees) to see if the behavior is a general one and not a special feature of the RIPPER algorithm. Finally we want to try this method on rare decays.

%\begin{thebibliography}{99}
%\bibitem{...} 
%....

\bibliography{proc}{}
\bibliographystyle{alpha}
  
%\end{thebibliography}

\end{document}